# The Helium content of Globular Clusters: light element abundance correlations and HB morphology.

## I. NGC 6752

S. Villanova,[1] G. Piotto,[2] R.G. Gratton[3]

[1] Grupo de Astronomia, Departamento de Fisica, Casilla 160, Universidad de Concepcion, Chile
    e-mail: `svillanova@astro-udec.cl`
[2] Dipartimento di Astronomia, Università di Padova, Vicolo dell'Osservatorio 3, I-35122 Padua, Italy
    e-mail: `giampaolo.piotto@unipd.it`
[3] INAF-Osservatorio Astronomico di Padova, Vicolo dell'Osservatorio 5, I-35122 Padua, Italy
    e-mail: `raffaele.gratton@oapd.inaf.it`

**ABSTRACT**

*Context.* In the context of the multiple stellar population scenario in globular clusters (GC), helium has been proposed as the key element to interpret the observed multiple main sequences (MS), subgiant branches (SGB) and red giant branches (RGB), as well as the complex horizontal branch (HB) morphology. However, up to now, He was never directly measured in GC stars (8500<$T_{eff}$<11500 K) with the purpose of verify this hypothesis.
*Aims.* We studied hot blue horizontal branch (BHB) stars in the GC NGC 6752 with the purpose to measure their Helium content. Our goal is to verify the feasibility of He measurement from high resolution spectra in stars cooler than 11500 K, where chemical abundances are not altered by sedimentation or levitation.
*Methods.* We observed 7 BHB stars using UVES@VLT2 spectroscopic facility. Spectra of S/N~200 were obtained and the very weak He line at 5875 Å measured. We compared this feature with synthetic spectra to obtain He abundances. In addition, iron peak (Fe,Cr), α (Si,Ti), light (O,Na), and s-element (Ba) abundances were measured.
*Results.* We could measure He abundance only for stars warmer than $T_{eff}$=8500 K. All our targets with measurable He are zero age HB (ZAHB) objects and turned out to have a homogeneous He content with a mean value of Y=0.245±0.012, compatible with the most recent measurements of the primordial He content of the Universe (Y~0.25). The whole sample of stars have a metallicity of [Fe/H]=-1.56±0.03 and [α/Fe]=+0.21±0.03, in agreement with other studies available in literature. Our HB targets show the same Na-O anticorrelation identified among the TO-SGB-RGB stars.
*Conclusions.* This is the first direct measurement of the He abundance for a significative sample of GC stars in a temperature regime where the He content is not altered by sedimentation processing or extreme mixing as suggested for the hottest, late helium flasher HB stars.

## 1. Introduction

In the last few years the argument of He content of Globular Clusters (hereafter GC) rose up thanks to the discovery of multiple sequences in some of these objects. The most interesting and peculiar cluster is ω Centauri, where at least 3 main sequences (hereafter MS) are present (Bedin et al. 2004; Villanova et al. 2007). The other very peculiar cluster showing 3 MS is NGC 2808 (Piotto et al. 2007). These sequences cannot be explained as a metallicity effect, because in ω Centauri Piotto et al. (2005) showed that the bluest MS is more metal rich than the main red population, while in NGC 2808 they all have the same iron content (as inferred from abundances in RGB stars). The only remaining parameter affecting the position of a star in the MS is the Helium content, and this was the explanation proposed for the interpretation of the photometric and spectroscopic properties of the MS stars in ω Centauri (Norris 2004; Piotto et al. 2005) and NGC 2808 (D'Antona & Ventura 2007; Piotto et al. 2007). In both cases, a He enhancement up to Y=0.40 is required.

The most natural explanation for the multiple sequence phenomenon is that these clusters had two or more episodes of star formation, where the younger populations were born from interstellar medium polluted by products of CNO cycle coming from massive stars of the first generation. In this picture the interstellar medium is affected by an enhancement of its He content, together with N, Na, and Al, while C and O turn out to be depleted. This hypothesis can explain also correlations or anticorrelations of light elements (i.e. C vs. N or Na vs. O, Marino et al. 2008) present also at the level of unevolved stars (Gratton et al. 2001). On the other hand this phenomenon cannot be explained by evolutionary effects like deep mixing precesses that happen in the RGB only after the first dredge-up or at the RGB-bump phase. Carretta et al. (2007) found also that the extention of the Na-O anticorrelation is related to the morphology of the HB, suggesting that this is an observational evidence that the anticorrelation may be connected also with the He spread.

In fact, it has been clear since the '60s that the HB morphology is related not only to the cluster iron content (as expected from

*Send offprint requests to*:

the models), but also to one or more other cluster parameters (the so called *second parameter problem*) which must account for the fact that some clusters have HB extended or extremely extended to the blue. Several explanations were suggested, including stellar rotation, age, cluster central density, mass (see Fusi Pecci et al. 1993 for an extended discussion), even infalling planets (Soker & Harpaz 2000). In this context it is interesting to note that also a spread in He abundance (and mass loss efficiency along the red giant branch which increase for He-rich stars) can reproduce the HB morphology in GCs (as first noticed by Rood 1973)

According to this picture, both stars with normal and enhanced He content end up on the zero-age horizontal branch (ZAHB) after the onset of core He burning, but He-rich stars should have warmer effective temperatures than He-normal ones. So He enhancement (and the related enhanced mass loss) could provide the needed mechanism required to spread stars from the redder and cooler HB (stars with normal composition) to the hotter and bluer part of the HB (He-rich stars) as discussed in D'Antona et al. (2002).

However, up to now this hypothesis was never confirmed, because a direct measurement of the He content for this purpose has never been attempted on GC stars, so far. The only targets available for this aim are the hot blue HB stars (hereafter BHB), belonging to the blue tail. As a matter of fact these stars (with $T_{eff} > 8000$ K) are sufficiently hot to show features of He in their spectra. On the other hand, only stars cooler than 11500 K are good candidates for a meaningful measurement of He because hotter ones show evidence of metal levitation (Behr 2005; Fabbian et al. 2005; Pace et al. 2006) and He sedimentation (Moni Bidin et al. 2007), which produces overabundance of heavy elements in the photosphere and a depletion of He, altering the original surface abundances. For this reason only a small window in $T_{eff}$ is available in the HB (and actually over the whole color magnitude diagram) of GCs to measure the He content. The aim of this paper is to verify if our purpose of measuring the He content in GCs is feasible observing BHB stars of a cluster with an extended blue HB tail as NGC 6752. This cluster was deeply studied both from photometric (Momany et al. 2002; Sabbi et al. 2004) and spectroscopic (Gratton et al. 2005; Yong et al. 2006) point of view, it has a well extended HB (including BHB) and O-Na anticorrelation, which are possibly connected with a spread in He abundance. For this reason it is the ideal target for our purposes. In Section 2 we describe the observations. In Sec. 3 and 4 we discuss the determination of the atmospheric parameters of our stars and the line list we used. Finally in Sec. 5, 6, 7 and 8 we present the results of the spectroscopic analysis, and briefly discuss and summarize our results.

## 2. Observations, Data Reduction and Membership

Observations of stars in the field of view of NGC 6752 were carried out in June-October 2007 in the context of the ESO Program ID 079.D-0674. In this program we observed a sample of 7 BHB stars (see Fig.1) for a total of 12×45 min. exposures, selected from B,V photometry obtained by the WFI imager at the ESO2.2m telescope.

The selected targets have spectral type A0 ($(B - V)_0 \sim 0.0$, $T_{eff} \sim 9000$ K). This choice allow us to have targets showing He features in their spectrum, but not yet affected by levitation. Observations were performed with the FLAMES-UVES spectrograph at the VLT@UT2(Kueyen) telescope. Spectra of the candidate BHB stars were obtained using the 580nm setting with 1.0" fibers. Spectra cover the wavelength range 4800-6800 Å

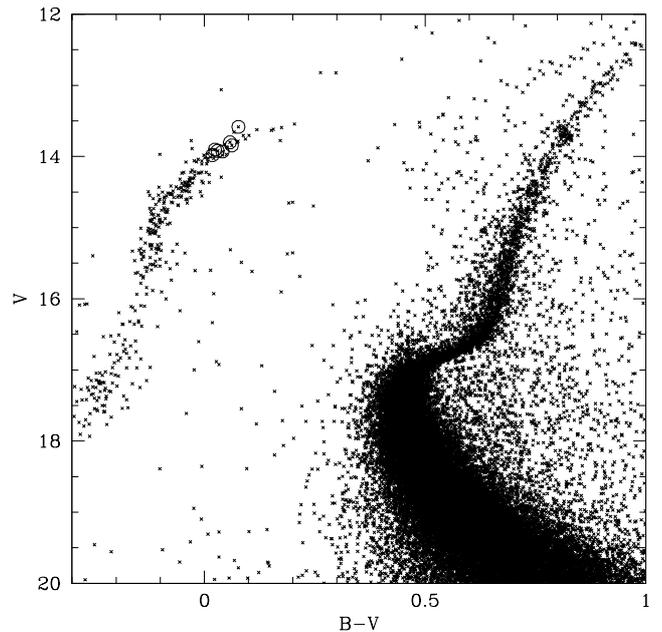

**Fig. 1.** The CMD of NGC 6752 with the observed BHB stars indicated as open circles.

with a mean resolution of R=47000.

Data were reduced using the UVES pipeline (Ballester et al. 2000), where raw data were bias-subtracted, flat-field corrected, extracted and wavelength calibrated. Echelle orders were flux calibrated using the master response curve of the instrument. Finally orders were merged to obtain a 1D spectrum and the spectra of each star averaged. Each spectrum have a mean S/N~200 per resolution element (high S/N and high spectral resolution are crucial for the success of this program, because the He lines are expected to be very weak).

The membership was established by radial velocity measurement. We used the *fxcor* IRAF utility to measure radial velocities. This routine cross-correlates the observed spectrum with a template having known radial velocity. As a template we used a synthetic spectrum calculated for a typical A0III star ($T_{eff}$=8500, $\log(g)$=3.00, $v_t$=1.00 km/s, [Fe/H]=-1.5, roughly the same parameters of our stars). Spectra were calculated using the 2.73 version of SPECTRUM, the local thermodynamical equilibrium spectral synthesis program freely distributed by Richard O. Gray[1].

The error in radial velocity - derived from *fxcor* routine - is less than 1 km/s. Finally, for the abundance analysis, each spectrum was shifted to rest-frame velocity and continuum-normalized.

Table 1 lists the basic parameters of the selected stars: the ID, the coordinates (RA & DEC), V magnitude, B-V and U-V colors, heliocentric radial velocity ($RV_H$), $T_{eff}$, $\log(g)$, microturbulence velocity ($v_t$), and projected rotational velocity (vsini) (for determination of atmospheric parameters and projected rotational velocity see the Sections 3 and 4). From the measured radial velocities we obtained a mean heliocentric radial velocity and a dispersion of:

$$< RV_H > = -27.2 \pm 1.8 \text{ km/s}, \sigma_{RV} = 4.7 \pm 1.3 \text{ km/s}$$

---

[1] See http://www.phys.appstate.edu/spectrum/spectrum.html for more details.

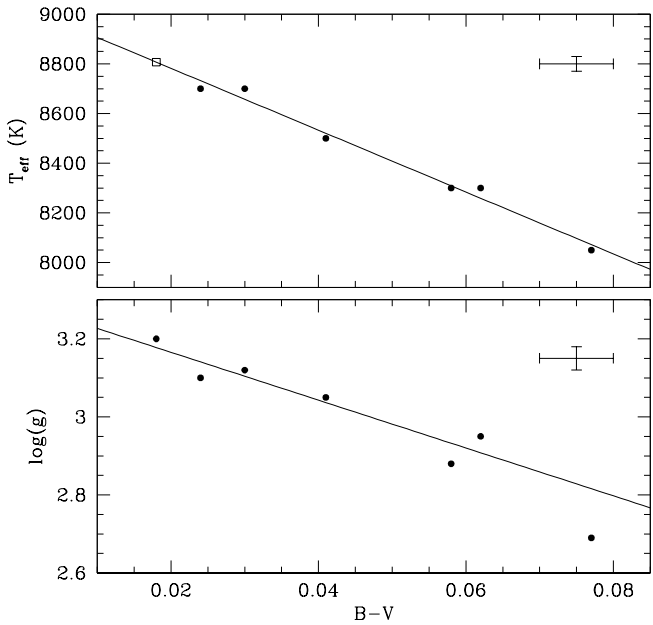

**Fig. 2.** $T_{eff}$ and log(g) values as a function of the B-V color for our sample of stars (filled circles). The open square is star #24938 for which no FeI and FeII lines were available for $T_{eff}$ determination. For this reason, its $T_{eff}$ was obtained from the extrapolation of these relations at the color of the star as explained in the text.

The mean velocity agree with literature values (see e.g. Harris (1996):$< RV_H > = -27.9$ km/s) and a dispersion of ∼4-5 km/s is typical for a massive cluster like NGC 6752 (Pryor & Meylan 1993). All our targets have radial velocity within 3 $\sigma$ with respect the mean value. Considering this and that at their position on the CMD there is not background contamination, we conclude that all the observed stars are members.

## 3. Atmospheric Parameters

We used the abundances from FeI/II and the shape of $H_{\alpha,\beta}$ features to obtain atmospheric parameters using the equivalent width (EQW) and spectral synthesis methods respectively.
All our stars show no strong line broadening due to rotation (vsini<10 km/s, see 1). For this reason EQWs would be obtained from a Gaussian fit of the spectral features.
We could measure only a small number of Fe lines for each spectrum (∼10-15 FeI lines and ∼8-10 FeII lines) due to the limited spectral coverage. However the high quality of our spectra (allowing an accurate measurement of the EQWs) and the simultaneous use of both FeI/II sets of lines allowed us to obtain reliable atmospheric parameters, as confirmed by the comparison with other works (see below). The analysis, with the exception of Hydrogen line calculation, was performed using the 2007 version of MOOG (Sneden 1973) under LTE approximation coupled with ATLAS9 model atmospheres (Kurucz 1992).

$T_{eff}$ was obtained by eliminating any trend in the relation of the abundances obtained from Fe I and Fe II lines with respect to the excitation potential, while microturbulence velocity ($v_t$) was obtained by eliminating any slope of the abundances obtained from FeI and FeII lines vs. reduced EQWs. On the other hand, log(g) values were estimated from comparison of the observed shape of $H_{\alpha,\beta}$ lines with synthetic spectra. In the $T_{eff}$ range of our stars the Hydrogen lines are not very sensitive to the temperature, so they can be used only for the gravity. Synthetic spectra for Hydrogen lines were calculated for the given temperature of each star and a wide range of gravities (2.0-4.0 dex) using the program SYNTHE [2] coupled with ATLAS9 model atmosphere. The log(g) parameter was obtained by the $\chi^2$ method.

Adopted values for the atmospheric parameters are reported in Table 1.
For the star #24938 (the warmest in our sample) we could not measure suitable Fe I or Fe II lines for $T_{eff}$ determination. For this reason we were forced to use the following procedure to obtain its value. Firstly, we obtained the empirical relations $T_{eff}$ vs. B-V color using the remaining stars. The adopted relation is the line plotted in Fig.2 (upper panel), which represents the best fit relation. Then, we extrapolated the relations to the color of #24938. Any extrapolation procedure can be very dangerous, but in this case the relation is very well defined and the amount of the extrapolation is small (0.006 mag in color) with respect the total interval covered by the empirical relation itself (0.054 mag). The microturbulence velocity was obtained in the usual way from the FeII lines. In the lower panel of Fig. 2, we plot log(g) vs. B-V color: also in this case a nice relation appears.
Interestingly, star #11583 (the reddest one) is well below the best fitting line. This is not surprising because, according to its position in the CMD, it is an evolved HB object (∼0.25 mag above the unevolved stars), so its gravity should be lower than for a ZAHB star, as we find from our analysis. This star has been excluded in deriving the best fitting line. We note that the offset of the gravity of this star from the fit line derived from the remaining objects (∼0.1 dex) corresponds exactly to the observed difference of ∼0.25 mag with respect to the ZAHB stars, assuming a similar mass.
All the other stars, according to their position on the CMD, are ZAHB objects.
From the relations obtained in Fig.2, we can obtain the typical internal error for our parameters. From the difference of the single points with respect the empirical relations we estimated mean errors for $T_{eff}$ and log(g) of 30 K and 0.05 dex, respectively. These errors are plotted as error bars in Fig. 2, coupled with the typical relative error in B-V color (0.005 mag). These errors are to be considered as internal. Systematic ones are likely much larger and more difficult to quantify.
The typical error in microturbulence can be obtained changing $v_t$ until the 1$\sigma$ value from the original slope of the relation between line strengths and abundances was reached. The corresponding internal error in microturbulence velocities is 0.04 km/s.
We tried to verify whether our $T_{eff}$ scale is correct comparing the observed colors with synthetic B,V photometry both from Kurucz [3] and Padova isochrone database (Girardi et al. 2002). For this purpose we assumed the E(B-V) value from Harris (1996). We obtained unclear results: while from Kurucz's data we obtained a good agreement only for cooler stars, from Padova isochrones the colors agree only for hotter ones. To solve the question we tried to apply the two $T_{eff}$ corrections as suggested by the two datasets separately, but in both cases we obtained a mean metallicity clearly not in agreement with literature values (see Sec. 5), reaching a difference of 0.3-0.4 dex. However, because of the fact that our final mean metallicity well agrees with literature values (see Sec. 5), we conclude that our $T_{eff}$ scale is not affected by strong systematic errors.

---

[2] See http://wwwuser.oat.ts.astro.it/castelli/
[3] http://kurucz.harvard.edu/

**Table 1.** Basic parameters for the observed stars.

| ID | RA(°) | DEC(°) | V(mag) | B-V(mag) | U-V(mag) | $RV_H$(km/s) | $T_{eff}$(K) | log(g)(dex) | $v_t$(km/s) | vsini (km/s) |
|---|---|---|---|---|---|---|---|---|---|---|
| 11583 | 287.724375 | -59.993889 | 13.584 | 0.077 | 0.152 | -23.8 | 8050 | 2.69 | 1.11 | 5 |
| 12404 | 287.738000 | -59.989694 | 13.903 | 0.024 | 0.081 | -31.6 | 8700 | 3.10 | 1.08 | 6 |
| 13415 | 287.667958 | -59.984639 | 13.800 | 0.058 | 0.137 | -30.6 | 8300 | 2.88 | 1.09 | 4 |
| 16149 | 287.793708 | -59.970889 | 13.844 | 0.062 | 0.124 | -20.4 | 8300 | 2.95 | 1.12 | 7 |
| 18646 | 287.729625 | -59.956361 | 13.920 | 0.030 | 0.084 | -23.9 | 8700 | 3.12 | 1.10 | 4 |
| 8745 | 287.676417 | -60.009639 | 13.931 | 0.041 | 0.100 | -26.8 | 8500 | 3.05 | 1.11 | 5 |
| 24938 | 287.670292 | -59.871917 | 13.983 | 0.018 | 0.035 | -33.1 | 8800 | 3.20 | 0.60 | 7 |

The last check on our $T_{eff}$ and log(g) scales was done comparing our results with Moni Bidin et al. (2007). In that paper, authors measure $T_{eff}$ and log(g) for a sample of HB stars using a different method and covering also the range of our targets. Unfortunately, there are no stars in common. However, we plotted both datasets against color and magnitude in Fig. 3, where our data are plotted as filled circles, while Moni Bidin et al. (2007) points are shown as open squares. The upper panels of Fig. 3 show $T_{eff}$ vs U-V color (right), and log(g) vs V magnitude (left). The lines are the best fitting 2$^{nd}$ order polynomial to the open squares. Present paper data nicely follow the fitted relations. A more quantitative comparison can be done calculating the mean difference between our data and the best fitting empirical relations. We found a difference in temperature of only ∼100 K, while no offset in gravity was found, implying a good agreement between the two datasets. The final comparison is shown in Fig. 3, lower panel, where we plotted the stars in the log(g) vs log($T_{eff}$) plane, as usually done for HB stars. Also in this case the agreement between our data and the empirical relations defined by Moni Bidin et al. (2007) (the fitted line) is very good.

Our conclusion is that our $T_{eff}$ and log(g) scales are not affected by large systematic effects.

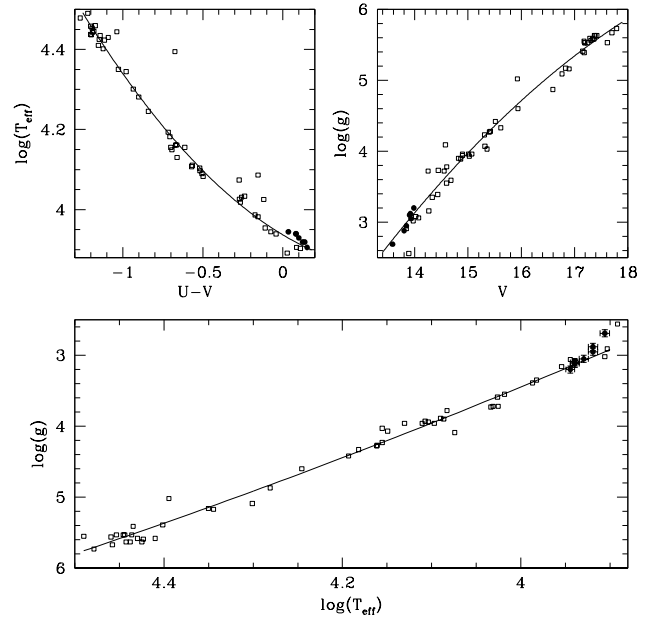

**Fig. 3.** Upper panels: log($T_{eff}$) vs U-V and log(g) vs V relations. Filled circles are our data, while open squares are data from Moni Bidin et al. (2007). Lines are the empirical relations as defined by open squares. Lower panel: the same data but in the log($T_{eff}$) vs log(g) plane.

## 4. Linelist, chemical abundances and rotation

For many elements (Na, Si, Ti, Cr, Fe, Ba), spectral lines are not blended, so abundances were obtained directly from EWQs. Spectral features affected by strong telluric contamination were rejected.

A suitable line list was taken from VALD database (Kupka et al. 2000) for a typical A0 star. log(gf) parameters were then redetermined by inverse analysis by measuring the equivalent widths on a reference spectrum: Vega spectrum was chosen for this purpose.

We measured EQWs on two high S/N Vega spectra, the first one (wavelength range 4100-6800 Å) obtained from the Elodie database (Moultaka et al. 2004), and the second one (wavelength range 3900-8800 Å) from Takeda et al. (2007). Vega is a rotating star **(vsini=21 km/s)** so the EQWs were obtained from direct integration of spectral features without fitting any curve (i.e. a Gaussian). In the common wavelength range (4100-6800 Å) the measured EQWs were averaged. The atmospheric parameters for Vega were obtained in the following way. As for the $T_{eff}$, Vega is primary standard because both angular diameter and bolometric flux are available. From literature we obtain: $f$=29.83±1.20×10$^{-9}$ W/m$^2$, $\theta$=3.223±0.008 mas. This turns out in a effective temperature of 9640±100 K. No direct estimate of Vega gravity is possible (i.e. independent from any kind of model), so we measured log(g) by fitting H Balmer lines. We obtained log(g)=3.97±0.05. Microturbulence velocity was obtained by minimizing the slope of the abundances obtained from FeI lines vs. reduced EQW. We obtained $v_t$=1.02±0.05 km/s and log$\epsilon$(Fe)=7.02 dex. The iron content referred to the Sun turns out to be [Fe/H]∼-0.5, in agreement with previous determinations (Qiu et al. 2001). Vega abundances we obtained from this line list (Na,Si,Ti,Cr,Fe,Ba) are reported in Tab. 2 and compared with Qiu et al. (2001) and with the Sun (Grevesse & Sauval 1998). Finally, for each element the log(gf) value of each line was adjusted in order to remove any difference in the derived abundance with respect the mean value obtained for that element. Vega turns out to have a mean iron content of [Fe/H]=-0.48±0.04.

At odd with the other elements for He and O we were forced to apply the spectral synthesis method because their spectral features at 5875 Å and 6156-6158 Å, respectively, are formed by several blended lines of the elements themselves. The line lists for the comparison with synthetic spectra were taken from VALD & NIST[4] databases and the log(gf) values simply averaged. Using spectral synthesis we could also measure projected rotational velocity for each star. For this purpose we assumed a combined instrumental+rotational profile for spectral features. Istrumental profile was assumed to be gaussian with FWHM=R/$\lambda$ (where R is the resolution of the instrument). Then vsini was varied in order to match the observed profile. Results are reported in Tab. 1. Typical error on vsini is 1-2 km/s, obtained from comparison of the results from He and O lines. An example of the spectral synthesis for these two elements is plotted in Fig.4 for #18646 star.

---
[4] See http://physics.nist.gov/PhysRefData/ASD/lines_form.html

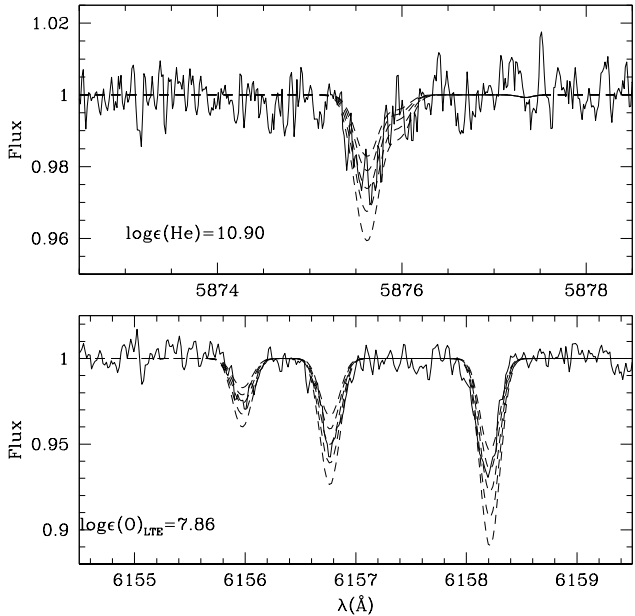
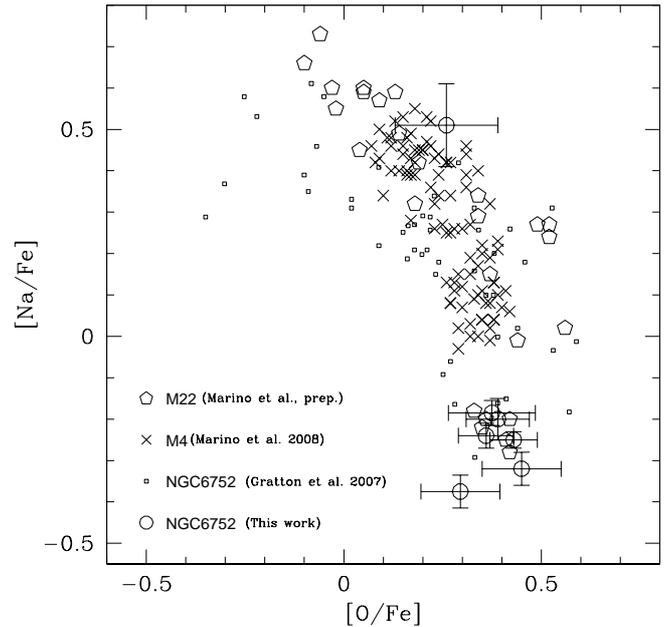

**Fig. 4.** Example of He (upper panel) and O (lower panel) spectral line fitting for the star #18646. Values for measured He and O abundances are indicated. Synthetic spectra with difference in abundance of -0.20,-0.10,+0.00,+0.10,+0.20 dex with respect the indicated value are shown.

**Fig. 5.** Na-O abundances found for our HB targets (Open circles). Open squares are data for TO-SGB-RGB stars in NGC 6752. Open pentagons are data for RGB stars in M22 while crosses are data for RGB stars in M4. See Sec. 5 for more details.

O abundance for Vega is reported in Tab. 2 while He abundance estimate was not possible because the target spectral line (5875 Å) was too contaminated by telluric features.

The analysis for Vega, as for the other stars, was performed using MOOG coupled with ATLAS9 model atmosphere.

The O triplet at 6156-6158 Å and the Na doublet at 5889-5895 Å are known to be affected by NLTE, and we applied the corrections by Takeda (1997) and Mashonkina et al. (2000) respectively, interpolated to the atmospheric parameters of our stars.

For Ba, whose lines are affected by hyperfine structure, no hyperfine data were available, so we did not apply any correction. The result can be an overestimation of its abundance. However our Ba lines are very weak (EQW<10 mÅ), so the final abundance is not affect by this problem because, according to McWilliam & Rich (1994), desaturation due to hyperfine structure is present only for EQW>50-60 mÅ.

## 5. Results of the spectroscopic analysis and determination of the metal content

The chemical abundances we obtained are summarized in Tab. 3. In this Table, abundances are expressed in log$\epsilon$(El)=log($N_{El}/N_H$)+12, where $N_{El}$ is the density of atoms of a given element. In the last two columns we give the average abundances and the average abundances relative to the Sun as obtained from the analyzed stars. In the calculation of the average abundances of O and Na we did not consider the star #11583 because its O and Na contents deviate from the general trend. This is due to the Na-O anticorrelation affecting the cluster as explained below.

From our sample the cluster turns out to have a Fe and $\alpha$-elements (as obtained from Si and Ti) content of:

$$[Fe/H] = -1.56 \pm 0.02, \quad [\alpha/Fe] = +0.21 \pm 0.04$$

(internal error only) which well agree with the results from Gratton et al. (2005) ([Fe/H]=-1.48±0.07 and [$\alpha$/Fe]=+0.27±0.01), with Harris (1996) ([Fe/H]=-1.56), and with Yong et al. (2006) ([Fe/H]=-1.61).

On the other hand both Cr and Ba appear to be slightly sub-solar ([Cr/Fe]=-0.08±0.04 and [Ba/Fe]=-0.15±0.09).

A more detailed discussion is required for the abundances of Na and O. According to Tab. 3 all but one target (#11583) appear to have roughly the same Na and O abundances within the errors. #11583 is significantly more Na-rich and more O-poor with respect the other stars. This behavior is expected because the cluster is affected by Na-O anticorrelation (Gratton et al. 2005). As said before, both elements were obtained from features known to be affected by non negligible NLTE deviation. NLTE corrections for O are well discussed in Takeda (1997), while the ones for Na are discussed in Mashonkina et al. (2000). Both papers present a grid of corrections which fully covers the parameter space of our objects. For each star we applied a NLTE correction interpolated from those grids using the atmospheric parameters obtained from our analysis. Correction for O turned out to be in the range -0.14÷-0.17 dex, while correction for Na is -0.20 for all stars. LTE and NLTE O and Na abundances are reported in Tab. 3. The corrections were applied also to O and Na abundance of Vega (see Tab. 2).

**Table 2.** Abundances for Vega as obtained from the line lists used in this work. Abundances for the Sun (Grevesse & Sauval 1998) and for Vega (Vega$_{Qi01}$) as obtained by Qiu et al. (2001) are reported. For some elements (O, Na) NLTE correction were applied and both LTE and NLTE abundances reported.

| El. | Vega | Vega$_{Qi01}$ | Sun |
|---|---|---|---|
| OI | 8.82 | 9.01 | 8.83 |
| OI$_{NLTE}$ | 8.74 | - | - |
| NaI | 6.63 | 6.45 | 6.33 |
| NaI$_{NLTE}$ | 6.33 | - | - |
| SiII | 7.17 | 6.96 | 7.55 |
| TiII | 4.52 | 4.58 | 5.02 |
| CrII | 5.20 | 5.19 | 5.67 |
| FeI | 7.03 | 6.94 | 7.50 |
| FeII | 7.01 | 6.93 | 7.50 |
| BaII | 1.85 | 0.81 | 2.13 |

**Table 3.** Abundances obtained for the target stars. In the two last columns we give the average abundances and the average abundances relative to the Sun. In the calculation of the average abundances of O and Na we rejected the star #11583 because its O and Na contents deviate from the general trend (see text). For #11583 we give also an upper limit for the He abundance, which is useful for the discussion, see Sec. 6. Errors for He abundance was obtained as described in Sec. 6.

| El. | 11583 | 12404 | 13415 | 16149 | 18646 | 8745 | 24938 | $<\log\epsilon(El)>$ | [El/H] |
|---|---|---|---|---|---|---|---|---|---|
| HeI | <11.1 | 10.89±0.07 | - | - | 10.90±0.07 | 10.84±0.07 | 10.99±0.07 | 10.91±0.03 | - |
| OI | 7.56±0.13 | 7.84±0.10 | 7.82±0.08 | 7.73±0.07 | 7.86±0.11 | 7.83±0.10 | 7.87±0.06 | - | - |
| OI$_{NLTE}$ | 7.41±0.13 | 7.70±0.10 | 7.68±0.08 | 7.56±0.07 | 7.72±0.11 | 7.68±0.10 | 7.70±0.06 | 7.67±0.02 | -1.16±0.03 |
| NaI | 5.35±0.10 | 4.72±0.04 | 4.78±0.05 | 4.65±0.03 | 4.85±0.03 | 4.60±0.04 | 4.71±0.02 | - | - |
| NaI$_{NLTE}$ | 5.15±0.10 | 4.52±0.04 | 4.58±0.05 | 4.45±0.03 | 4.65±0.03 | 4.40±0.04 | 4.51±0.02 | 4.52±0.04 | -1.81±0.04 |
| SiII | 6.17±0.13 | 6.31±0.10 | 6.24±0.10 | 6.20 | 6.28±0.12 | 6.25±0.09 | 6.18 | 6.23±0.02 | -1.32±0.02 |
| TiII | 3.61±0.05 | 3.74±0.04 | 3.64±0.03 | 3.58±0.03 | 3.71±0.03 | 3.60±0.04 | 3.60 | 3.64±0.02 | -1.38±0.02 |
| CrII | 3.92±0.05 | 4.17±0.06 | 4.06±0.02 | 3.96±0.04 | 4.00±0.02 | 3.98±0.04 | 4.12 | 4.03±0.03 | -1.64±0.03 |
| FeI | 5.83±0.02 | 6.14±0.02 | 5.97±0.02 | 5.82±0.03 | 6.08±0.02 | 5.91±0.02 | - | 5.96±0.05 | -1.54±0.05 |
| FeII | 5.81±0.04 | 6.01±0.05 | 5.95±0.07 | 5.92±0.06 | 5.95±0.05 | 5.89±0.05 | 5.94±0.04 | 5.92±0.02 | -1.58±0.02 |
| BaII | 0.25 | - | 0.45 | 0.55 | - | - | - | 0.42±0.09 | -1.71±0.09 |

We checked final O and Na abundance determination by comparing in Fig. 5 our results with the Na-O anticorrelations found for NGC 6752 by Gratton et al. (2005). We compared our results also with anticorrelation found for M 4 (Marino et al. 2008) and with the recent result for M 22 (Marino et al. 2009). There is a good match, within the errors, between our data and Gratton et al. (2005). The agreement is even better with anticorrelations found in M4 and M22. In this case our data nicely follow the shape of the anticorrelations defined by M 22 and M 4 RGB stars.

Our conclusion is that our O and Na determinations are not affected by significant systematic errors, and that O and Na content of #11583 can be fully explained by the Na-O anticorrelation found in NGC 6752 and in other clusters.

## 6. Helium content and the problem of levitation

Finally, and most importantly, we discuss the measurement of the He content of our targets, which was the main driving goal for this paper. We were actually able to measure reliable He abundances only in the warmer ZAHB stars of our sample ($T_{eff}$ >8500 K), the He feature being too weak in the cooler objects (however we were able to give an upper limit for #11583, which turns out to be important for the following discussion). In Fig. 6 we plotted He lines for these stars with the best fitting synthetic spectrum. This gives an idea of the quality of our data. This group of stars turns out to have a mean He content of $<\log\epsilon(He)>$ =10.91±0.03 with a dispersion $\sigma$=0.06±0.02 dex. This translates in:

$$Y = 0.245 \pm 0.012$$

We can compare this value with the primordial one. The most recent empirical work on the primordial He content of the Universe is by Izotov et al. (2007). In this paper the authors find a He content of Y=0.2472±0.0012 or Y=0.2516±0.0011, depending on the used method.

In order to verify if these stars have also an homogeneous He content, we performed a detailed analysis of internal error for this element. He was measured by comparison of the observed spectrum with 5 synthetic ones, adopting the value that minimizes the r.m.s. scatter of the differences. S/N of the real spectrum introduces an error in the final He abundance which can be estimated calculating the error on the r.m.s. scatter. For our data this error turns out to corresponds to a uncertainties in abundance of 0.07 dex. To this value me must add errors due to the uncertainties on atmospheric parameters. $\Delta T_{eff}$=30 K gives $\Delta$He=0.015 dex, $\Delta\log(g)$=0.05 gives $\Delta$He=0.020 dex, while the error on microturbulence has no appreciable influence on He abundance.

The final uncertainties in He abundance is given by the squared sum of all the single errors, and the final result is $\Delta$He$_{tot}$=0.07 dex. This shows that the main contribution to the final error comes from the S/N of the real spectrum. If we compare this value with the observed dispersion (0.06±0.02 dex), we can conclude that the ZAHB stars on the red side of the HB have a homogeneous He content which is compatible within the errors with the primordial value.

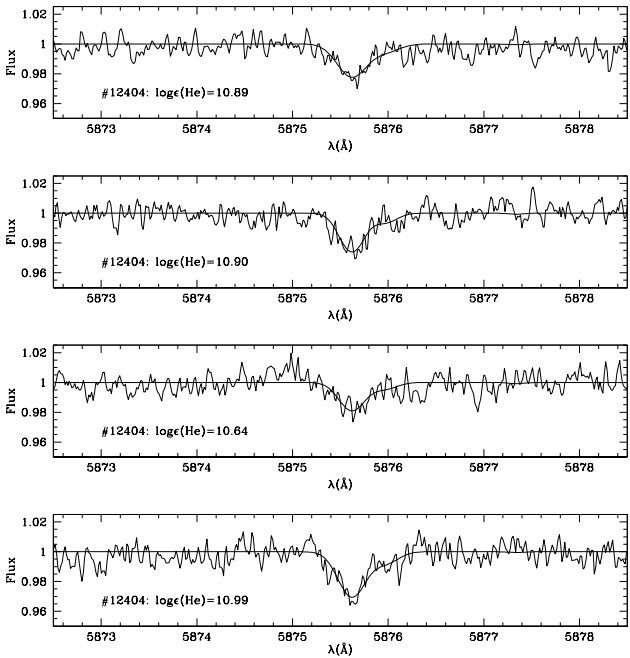

**Fig. 6.** He lines of the stars for which this element could be measured. We superimposed to each observed spectrum the best fitting synthetic one.

This fact suggest also we are measuring the He content that those stars had when they formed (i.e. the primordial value of the cluster). To further prove this assertion we should demonstrate that our object are not affected by levitation of heavy elements and He sedimentation.

As discussed before levitation and sedimentation are present for temperatures hotter than 11500 K. Being our stars cooler we expect that they are not affected by these phenomena. With the purpose of verifying this hypothesis we plotted $\log\epsilon$(He), [El./Fe] and [FeI-II/H] vs. $T_{eff}$ in Fig. 7.

In this figure, dashed lines represent the upper and lower abundance limits found by Gratton et al. (2005), Yong et al. (2006), and Yong et al. (2008) for RGB stars, when available.

For He we have too few points (four) in a too small $T_{eff}$ range, for a conclusive assessment of the absence of sedimentation.

On the other hand Si, Ti, Cr and FeII do not show any evidence of trend of overabundance to be associated to levitation, within the errors.

Also O and Na do not show evidence of any trend with the exception of the coolest star (#11583) which is slightly O-poorer (~-0.1 dex) and strongly Na enriched (by ~+0.7 dex) with respect to the other objects due to the Na-O anticorrelation.

Some suspect for a trend could be present for FeI, though it is at most confined to the two hotter stars. Their overabundance (~0.10÷0.15 dex) could be due to NLTE effect. However other stars, that are only slightly cooler (200 K), show no evidence of NLTE effects. In addition, an error of ~0.1 dex is typical for the abundance of a single stars, implying that the trend for FeI is due measurement errors.

Considering also that our abundances well agree with RGB values, we conclude from an observational point of view that the elements measured in the present paper do not show evidence of levitation, and consequently we can exclude also He sedimentation.

It is interesting to compare this last results with Michaud et al. (2008), which analyzes in detail from a theoretical point of view the change in surface abundances due to levitation.

As the authors point out theoretical models can well explain abundance anomalies of many elements (in particular Fe, see their Fig. 9) observed for HB stars hotter than 11500 K in M15. On the other hand models predict a strong Fe overabundance (up to +1.0 dex) also for objects cooler than 11500 K, which is not observed at all, neither in M15 nor in our case.

As suggested by the authors atmospheric mixing due to rotation (a vsini up to 35 km/s was observed in many HB stars) can be the explanation for this behavior.

However at odd with that we did not find any fast rotating stars in our sample, which is a pretty sure result because stars having vsini larger than 10 km/s can be easily recognized by the shape of their spectral line (as in the case of Vega for which we found a rotational velocity of 21 km/s).

We could assume that all our objects have a real rotational velocity much larger, and that the observed vsini value is so small due to the inclination, but the probability to pick up by chance 7 star having vsini<10 km/s as in our case is very low (<<1% assuming an uniform distribution of the inclination angle and a common value for the real rotational velocity of 35 km/s).

Our final conclusion is that chemical abundances measured in the present paper are not affected at all by levitation or sedimentation, as already found for other HB stars in the same $T_{eff}$ range (i.e. in M15). This fact is still not well understood because models predict abundance anomalies also for stars cooler than 11500 K, which are not observed, suggesting some mechanism other that rotation working against levitation. However some kind of connection between rotation and levitation seems to be indubitable, as found for example by Recio-Blanco et al. (2004).

## 7. Discussion

Correlations and anticorrelations of chemical elements observed in GCs (i.e. Na vs. O and Mg vs. Al) are attributed to contamination by products of the H-burning process at high temperature (Langer et al. 1993, Prantzos et al. 2007), when N is produced at the expenses of O, and proton capture on Ne and Mg produces Na and Al (CNO cycle). Gratton et al. (2001) demonstrated that this contamination actually does not occur along the RGB phase, but it is present also at the level of MS. This means that it is not the result of some mixing mechanism present when a star leaves the MS, but it is rather due to primordial pollution of the interstellar material from which stars were formed.

Pollution must come from more massive stars. GCs must have experience some chemical evolution at the beginning of their life (see Gratton et al. 2004 for extensive references). The main product of H-burning is He and an He enhancement is then expected to be present in stars with enhancement of N, Na, and Al. The favorite classes of candidate polluters are three: fast-rotating massive main-sequence stars (Decressin et al. 2007), intermediate-mass AGB stars (D'Antona et al. 2002), and primordial population III stars (Choi & Yi 2007). All these channels can potentially pollute the existing interstellar material with products of complete CNO burning, including He (see Renzini 2008 for an extensive review).

In the pollution scenario, a first generation of O-rich and Na-poor (with respect to a second stellar generation) stars is formed from primordial material, which, in case, must have been somehow polluted by previous SN explosions. This generation has also a primordial or close to primordial He content (Y=0.24-0.25). Then the most massive stars of this pristine stellar generation pollute the interstellar material with products of the CNO cycle. This material is kept in the cluster due to the relatively strong gravitational field, and it gives origin to a new generation of O-poor and Na-rich stars. This population should also have been He-enhanced (Y=0.28-0.35, depending on the favorite polluter). Also the abundance of other elements (including s-process elements) may differ in stars of the first or second generation.

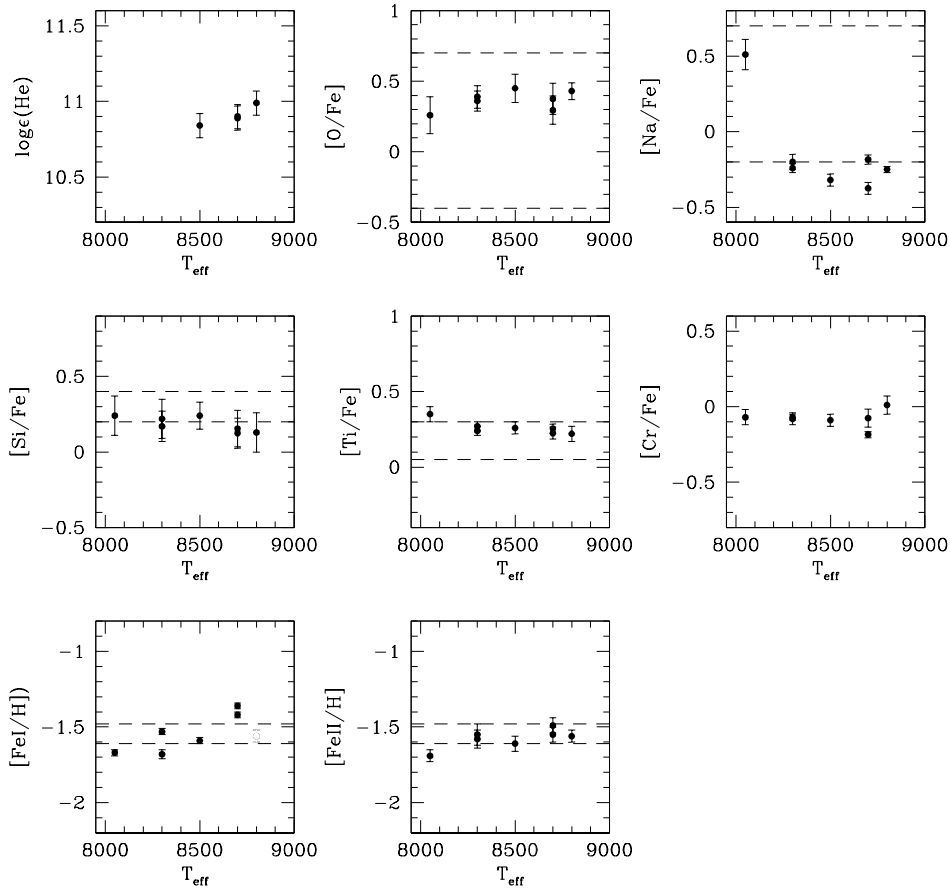

**Fig. 7.** Plot of log$\epsilon$(He), [El./Fe] and [FeI-II/H] vs. $T_{\rm eff}$. In this figure dashed lines represent the upper and lower abundance limits found by Gratton et al. (2005), Yong et al. (2006), and Yong et al. (2008) when available. See Sec. 5 for a detailed discussion of this figure.

In the MS phase He-rich stars evolve more rapidly than He-poor ones, so He/Na-rich (and O-poor) stars presently at the turn-off or in later phases of evolution should be less massive than He/Na-poor (and O-rich) ones. In this framework, D'Antona et al. (2002) and Carretta et al. (2007) proposed that a spread in He, combined with mass loss along the RGB, may be the ingredient to naturally reproduce the whole HB morphology in GCs, as discussed in the introduction. According to this scenario, primordial He, O-rich and Na-poor stars end up on the red, while stars with extreme abundance alterations (strong He enhancement, O-poor and Na-rich) may end up in the blue HB, if they have experienced enough mass loss during the RGB phase. In principle, the HB represents the ideal locus to investigate the effects of chemical anomalies in GC stars, as the HB is a sort of amplifier of the physical conditions that are the consequence of the star composition and previous evolution.

In this scenario, in NGC 6752, we expect that the progeny of RGB He-normal, O-rich and Na-poor stars should reside at the red part of the ZAHB. Therefore, it is not surprising that the observed ZAHB stars (all located on the cool side of the HB, at $T_{\rm eff} < 9,000$ K) are all O-rich, Na-poor, and have a close to primordial He abundance. Noteworthy, star #11583 is evolved off the ZAHB. At the color of our target HB stars, evolution off the ZAHB occurs quite parallel to the ZAHB itself, at brighter magnitude and toward redder color. This star is than expected to have been bluer when in the ZAHB. This means that there is a high probability that star #11583 may have been much bluer when it arrived in the HB, though it is not possible to infer its exact ZAHB position. It is then not surprising that it is also more Na-rich and O-poor than the other stars studied in this paper, and it could represent the progeny of RGB O-poor and Na-rich stars that we know to be present at earlier evolutionary phases. Because of this, it is also expected to be more He-rich. Unfortunately this star is too cool for any reliable derivation of the He-abundance. However we could establish an upper limit for its He abundance which turned out to be log$\epsilon$(He)<11.1 (Y<0.33), which is compatible with our hypothesis. New, properly tuned observations are mandatory to confirm the presence in HB of the He enhanced progeny of RGB O-poor and Na-rich stars.

## 8. Summary

We studied a sample of HB stars with a temperature in the range 8000-9000 K, with the aim of measuring their He content. Targets were selected in order to be hot enough to show the He feature at 5875 Å, but cold enough to avoid the problem of He sedimentation and metal levitation affecting HB stars hotter than 11500 K. Thanks to the high resolution and high S/N of our spectra, we were able to measure He abundances for stars warmer than 8500 K. Noteworthy, this is the first, direct measurement of He content from high resolution spectra of GC stars in this $T_{eff}$ regime.

Our sample of stars turns out to have [Fe/H]=-1.56±0.03 and [$\alpha$/Fe]=+0.21±0.03, values that well agree with the literature ones. This confirms that our study is not affected by strong systematic effects.

Our targets show the same Na-O anticorrelation already found in the TO-SGB-RGB region. We note that all our unevolved ZAHB targets are O-rich and Na-poor, and have a primordial He content (Y=0.254±0.012). This is what we expect if their position on the HB is driven by their He and metal content. The only O-poor, Na-rich star in our sample is an evolved object, that likely occupied a much bluer location when in the ZAHB. Interestingly enough, it has the highest Na content and the lowest O content among our target stars. According to the pollution scenario, we expect that it also has a He content higher than the primordial value. Unfortunately, because of its temperature, we cannot measure its He abundance.

Our work demonstrates the feasibility of measuring the He content in GC stars cold enough to show no sedimentation or levitation phenomena. The results of the present pilot project discloses new research possibilities. The first one is the direct measurement of the He content in clusters where light element correlation and anticorrelation were observed. Abundance of these element should be correlated also with He content, and now we have a way to directly prove this hypothesis. A second noteworthy project is the study of more complex objects, like $\omega$ Centauri and NGC 2808, where strong He enhancement has been proposed in the literature in order to explain the photometric and spectroscopic properties of their peculiar, multiple stellar populations.

*Acknowledgements.* We wish to warmly thank the referee for the careful reading of the manuscript and for her/his suggestions which surely helped us to improve the presentation of our results. GP and RG acknowledge partial support by MIUR under the program PRIN2007 (prot. 20075TP5K9).